\documentclass[12pt]{article}

\usepackage{amssymb}
\usepackage{cite}

\makeatletter
\renewcommand\thesubsection{\thesection.\@arabic\c@subsection}
\makeatother

\newcommand{\sect}[1]{\setcounter{equation}{0}\section{#1}}

\newcommand {\beq}{\begin{equation}}
\newcommand {\eeq}{\end{equation}}
\newcommand {\beqa}{\begin{eqnarray}}
\newcommand {\eeqa}{\end{eqnarray}}         
\newcommand {\beqs}{\begin{eqnarray*}}
\newcommand {\eeqs}{\end{eqnarray*}}
\newcommand {\bds}{\begin{displaymath}}
\newcommand {\eds}{\end{displaymath}}
\newcommand {\n}{\nonumber\\}

\newcommand {\bebb}{\begin{thebibliography}}
\newcommand {\eebb}{\end{thebibliography}}      
\newcommand {\bbit}{\bibitem}

\def\a{\alpha}

\def\k{\kappa}

\def\l{\lambda}

\def\pd{\prod}

\def\ra{\rangle}

\def\dg{\dagger}

\def\journal#1&#2(#3){\unskip, \sl #1\ \bf #2 \rm(19#3) }
\def\andjournal#1&#2(#3){\sl #1~\bf #2 \rm (19#3) }

\def\dz{\frac{d}{dz}}

\begin{document}


\begin{flushright}
\end{flushright}

\vskip 1cm

\begin{center}
{\Large\bf Exact solutions for a family of spin-boson systems}

\vspace{1cm}

{\large Yuan-Harng Lee, Jon Links, and Yao-Zhong Zhang}
\vskip.1in

{\em School of Mathematics and Physics,
The University of Queensland, Brisbane, Qld 4072, Australia}


\end{center}

\date{}



\begin{abstract}

We obtain the exact solutions for a family of spin-boson systems. This is achieved through application of the 
representation theory for polynomial deformations of the $su(2)$ Lie algebra. 
We demonstrate that the family of Hamiltonians includes, as special cases, known physical models 
which are the two-site Bose-Hubbard model, the Lipkin-Meshkov-Glick model, the molecular asymmetric rigid rotor, 
the Tavis-Cummings model, and a two-mode generalisation of the Tavis-Cummings model.

\end{abstract}

\vskip.1in

{\it PACS numbers}:  02.30.Ik; 03.65.Fd.

{\it Keywords}: exactly solvable models, Bethe ansatz.



\setcounter{section}{0}
\setcounter{equation}{0}

\sect{Introduction }

The study of polynomial deformations of Lie algebras is an area of research which has found many applications in systems involving non-linear interactions \cite{Karassiov94,Karassiov92-00,Karassiov02}. In recent publications \cite{Yuan10} we have formulated such methods for the analysis of a class of multi-boson systems. The approach of \cite{Yuan10} is to express the Hamiltonian of the systems in terms of the generators of polynomial deformations of the 
$su(2)$ Lie algebra, through the explicit construction of Fock-space representations. By utilising a correspondence between the Fock-space representations and differential operator realizations, it was shown that exact solutions are obtained in terms of a system of coupled equations. These equations can be viewed as providing a Bethe ansatz type of solution for the calculation of the energy spectrum and associated eigenstates. The generality of this approach allows for application on a wider level. The work described below is concerned with extending these methods to the study of a family of Hamiltonians which couple multi-boson degrees of freedom to a spin degree of freedom. In this manner we unify the problem of exactly solving spin-boson Hamiltonians to a particular class which contains within it a number of models which are already known in the literature, as we will discuss.             

The main result of this paper is the derivation of the exact eigenfunctions and energy
eigenvalues of the infinite family of spin-boson systems defined by the Hamiltonian
\beqa
H&=&\sum_{i=1}^{M}w_{i}N_i+ g'J_0^{s} + g\left(J_+^ra_{1}^{k_{1}}\cdots a_{r}^{k_{M}} 
+J_-^ra_1^{\dagger k_1}\cdots a_r^{\dagger k_M}\right),  \label{bosonH1}  
\eeqa
where throughout $r,s\in{\mathbb Z}_+$, $M, k_1,\cdots, k_M\in {\mathbb N}$,  $a_i,~a_i^\dagger$ and $N_i=a^\dagger_ia_i$
are bosonic annihilation, creation, and number operators respectively, $J_{\pm,0}$ are the generators for the $su(2)$ spin algebra, 
and $w_i,~w_{ij},~g$ are real coupling constants. 
The Hamiltonian of the form (\ref{bosonH1}) appears in the description of various physical systems of interest in atomic, 
molecular, nuclear and optical physics. We will explicitly demonstrate that (\ref{bosonH1}) includes as special cases several known models which are the 
two-site Bose-Hubbard model \cite{Leggett01}, the Lipkin-Meshkov-Glick model \cite{Lipkin65}, 
the molecular asymmetric rigid rotor \cite{King43}, the many-atom Tavis-Cummings model \cite{Tavis68,Hepp73} and a two-mode generalized Tavis-Cummings model \cite{Rybin98}.  
    
This paper is organized as follows. In section 2 we introduce new higher order polynomial algebras which are
dynamical symmetry algebras of Hamiltonian (\ref{bosonH1}), which enable an algebraization of the spin-boson systems.
We construct finite-dimensional unitary representations of the dynamical symmetry algebras in section 3 and the 
corresponding single-variable differential operator realizations in section 4. This leads to the higher order differential operator realizations 
of the Hamiltonian (\ref{bosonH1}). 
In section 5 we establish the quasi-exact solvability of this differential operator \cite{Turbiner88,Ushveridze94,Gonzarez93}
and solve for the eigenvalue problem via the functional Bethe ansatz method (see e.g. \cite{Wiegmann94,Sasaki09,Sasaki01}). 
In section 6 we present explicit results for several special cases,
thus providing a unified derivation of exact solutions to the widely-studied models mentioned above.  
We summarize our results in section 7 and discuss further avenues for investigation.



\sect{Algebraization} 

In this section we introduce new higher order polynomial deformations of $sl(2)$ and give an algebraization of 
the Hamiltonian (\ref{bosonH1}). Our approach extends previous studies 
\cite{Karassiov94,Karassiov92-00,Karassiov02,Yuan10} where more restricted classes of 
systems have been exactly solved using polynomial algebra structures. 

We introduce generators
\beq 
{\cal P}_+ = P_+ \pd_{i=1}^M Q_-^{(i)} ~,~~~ {\cal P}_- = P_- \pd_{i=1}^M Q_+^{(i)} ~,~~~  {\cal P}_0 = \frac{P_0-\sum_{i=1}^MQ_0^{(i)}}{M+1},
 \label{Jordan-1} 
\eeq 
where 
\begin{eqnarray*} 
&&P_0 = \frac{J_0}{r} , ~~~~ P_+ =J_+^r, ~~~~P_-=J_-^r, \n
&&Q_+^{(i)}= \frac{a_i^{\dg k_i}}{\sqrt{k_i}^{k_i}}~,~~~~ Q_-^{(i)} = \frac{a_i^{k_i}}{\sqrt{k_i}^{k_i}}~,~~~~ 
    Q_0^{(i)} = \frac{1}{k_i}\left(a_i^\dg a_i+\frac{1}{k_i}\right) 
\end{eqnarray*}
are $M+1$ mutually commuting operators. It can be shown that ${\cal P}_{\pm,0}$ satisfy the following commutation relations:
\beqa 
\left[{\cal P}_0,{\cal P}_{\pm} \right] &=& \pm {\cal P}_{\pm}  \n 
\left[{\cal P}_+,{\cal P}_- \right] &=& \psi^{(2r)}\left({\cal K},{\cal P}_0-1, C \right)
  \pd_{i=1}^M\phi^{(k_i)}\left({\cal K},{\cal P}_0-1,  \{{\cal L}\} \right) \n 
&-& \psi^{(2r)}\left({\cal K},{\cal P}_0, C \right)\pd_{i=1}^M\phi^{(k_i)}\left({\cal K},{\cal P}_0,  
  \{{\cal L}\} \right),\label{plsu2}
\eeqa
where $C$ is the Casimir operator of $su(2)$,
\beqa 
{\cal K}= \frac{MP_0+\sum_{\nu=1}^MQ_0^{(\nu)}}{M+1}~~,~~{\cal L}_i = Q_0^{(i)}-Q_0^{(i+1)}, ~~ i=1 \cdots M-1, \label{Jordan-C}
\eeqa 
are $M$ central elements of (\ref{plsu2}) and  
\begin{eqnarray*}
\psi^{(2r)}\left({\cal K},{\cal P}_0, C \right) &=&  -\pd_{i=1}^r\left[C-(r{\cal K}+r{\cal P}_0+r-i+1)
     (r{\cal K}+r{\cal P}_0+r-i)\right],  \\ 
\phi^{(k_i)}\left({\cal K},{\cal P}_0,  \{{\cal L}\} \right) &=& - \pd_{i=1}^{k_i}\left(\frac{{\cal K}}{M}-({\cal P}_0+1)
      -\frac{1}{M}\sum_{\mu=1}^{M-1}\mu{\cal L}_{\mu}+\sum_{\mu=i}^{M-1}{\cal L}_{\mu} +\frac{ik_i-1}{k_i^2} \right) 
\end{eqnarray*}
are polynomial functions of degree $2r$ and $k_i$ respectively. Thus (\ref{plsu2}) defines a polynomial algebra of degree $2r+\sum_{i=1}^Mk_i -1$.

In terms of the generators of the polynomial algebra (\ref{plsu2}), the Hamiltonian (\ref{bosonH1}) can be written as 
\beq
H=\sum_{i=1}^{M} w_iN_i + g'r^s \left({\cal P}_0+ {\cal K} \right)^{s} 
+ g\left[\pd_{i=1}^{M}\left(\sqrt{k_i}\right)^{k_i}\right]\left( {\cal P}_+ + {\cal P}_- \right)\label{bosonH2}
\eeq
with the number operators having the following expression in ${\cal P}_0$, ${\cal K}$ and ${\cal L}_i$
\begin{eqnarray*}  
N_i = k_i\left(-{\cal P}_0 +\frac{{\cal K}}{M}-\frac{1}{M}\sum_{\mu=1}^{M-1}\mu{\cal L}_\mu+\sum_{\mu=i}^{M-1}{\cal L}_\mu \right) -\frac{1}{k_i}.
\end{eqnarray*}
It follows that the polynomial algebra (\ref{plsu2}) is the dynamical symmetry algebra of the Hamiltonian (\ref{bosonH1}).

\sect{Unitary irreducible representations}

Irreducible representations of the polynomial algebra (\ref{plsu2}) can be constructed in the tensor product space of
the representation space of $P_{\pm,0}$ and  the Fock spaces of $\{Q_{\pm,0}^{(i)}\}$.
As shown in \cite{Yuan10}, the Fock states for irreducible representations of $\{Q_{\pm,0}^{(i)}\}$ 
are labelled by quantum numbers $q_i=\frac{1}{k_i^2},~\frac{k_i+1}{k_i^2},~\cdots,~ \frac{(k_i-1)k_i+1}{k_i^2}$,
through
\beqa 
|q_i,m_i \ra = {a^{\dg k_i(m_i+q_i-{k_i^{-2}})} \over \sqrt{[k_i(m_i+q_i-{k_i^{-2}})]!}}|0 \ra.~~~~
m_i &=& 0,1,\cdots.\label{fock-state1}
\eeqa 
The action of $Q_{\pm,0}^{(i)}$ on these states is
\begin{eqnarray*}
Q_0^{(i)}|q_i,m_i\ra &=& (q_i+m_i)|q_i,m_i\ra, \\
Q_+^{(i)}|q_i,m_i\ra &=& \pd_{j=1}^{k_i}\left(m_i+q_i+ \frac{j k_i-1}{k_i^2}\right)^{1/2}\,|q_i,m_i+1\ra, \\
Q_-^{(i)}|q_i,m_i\ra &=& \pd_{j=1}^{k_i}\left( m_i+q_i -\frac{(j-1)k_i+1}{k_i^2}\right)^{1/2}|q_i,m_i-1\ra.
\end{eqnarray*}
The irreducible representations of $P_{\pm,0}$  can be deduced from the $su(2)$-module $V_j,~j=0,\frac{1}{2},1,\cdots $ 
as follows. First, it can be shown that $P_{0,\pm}$ satisfy the relations 
\beqa 
\left[ P_0,P_{\pm} \right] &=& \pm P_{\pm}, \n 
\left[ P_+, P_-\right] &=& \psi^{(2r)}(P_0, C) - \psi^{(2r)}(P_0-1, C), \label{plsu2-P+-0}
\eeqa  
where 
\beqa 
\psi^{(2r)}\left(P_0, C \right) = -\pd_{i=1}^r\left(C-(rP_0+r-i+1)(rP_0+r-i)\right) \label{f1}
\eeqa 
is a polynomial in $P_0$ and $C$ of degree $2r$. Thus (\ref{plsu2-P+-0}) is a polynomial algebra of degree $2r-1$.
The Casimir operator of (\ref{plsu2-P+-0})
takes fixed value $\pd_{i=1}^r(C-i(i-1))$. 

It is easily verified that there are $\textrm{min}\{r ,2j+1\}$ lowest weight states, 
\begin{eqnarray*}
|j,0;p \ra &\sim &  J_+^p|j,0 \ra , ~~~~p=0,1,\cdots, \textrm{min}\{r-1 ,2j\},   
\end{eqnarray*}
where $|j,0 \ra$ is the lowest weight state of $su(2)$.  
This implies that finite-dimensional irreducible representations of (\ref{plsu2-P+-0}), denoted as $ V_{j,p}$,  are labelled by quantum numbers
$j$ and $p$, $j=0,\frac{1}{2},1,\cdots$, and $p=0,1,\cdots, \textrm{min}\{r-1 ,2j\}$. 
Thus we have the branching rule from $su(2)$ representation $V_j$ into $V_{j,p}$ of (\ref{plsu2-P+-0}):
\begin{eqnarray*} 
V_j = \oplus_{p=0}^{\textrm{\tiny min}\{(r-1),2j\}} V_{j,p}.
\end{eqnarray*}
General basis vectors in the irreducible representation space $V_{j,p}$ are given by $|j,n;p\ra \sim  (P_+)^n|j,0;p \ra $. Explicitly,
\beqa 
|j.n;p \ra = \sqrt{\frac{(2j-p-rn)!}{(p+rn)!(2j)!}}J_+^{p+rn}|j,0 \ra. \label{fock-state2}
\eeqa 
The action of $P_{0,\pm}$ on these vectors is given by 
\beqa 
P_0|j,n;p \ra  &=& \left(\frac{p-j}{r}+n\right) |j,n;p \ra, \n  
P_+|j,n;p \ra  &=& \pd_{i=1}^r \sqrt{(p+i+rn)(2j-p-i+1-rn)}\;|j,n+1;p \ra, \n 
P_-|j,n;p \ra  &=& \pd_{i=1}^r \sqrt{(p-i+1+rn)(2j-p+i-rn)}\;|j,n-1;p\ra. \label{plsu2rep}
\eeqa 
It can also be shown that 
\begin{eqnarray*} 
P_-|j,0;p\ra=0,~~~~~P_+|j,\frac{2j-p-\l}{r};p \ra  = 0, 
\end{eqnarray*}
where $\l$ is a non negative integer taking specific values $\l =0,1,\cdots, \textrm{min}\{r-1,2j \} $ according to $j$ and $p$. 
Moreover, $\frac{2j-p-\l}{r}$ is always a non-negative integer. Therefore $ n = 0,1,\cdots, \frac{2j-p-\l}{r}$, 
and (\ref{plsu2rep}) is a finite-dimensional representation of (\ref{plsu2-P+-0}) with dimension  $\frac{2j-p-\l}{r}+1$. 



We now construct irreducible representation of (\ref{plsu2}) in the tensor space $V_{j,p}\otimes {\cal H}^{(1)}_{q_1}\cdots\otimes {\cal H}^{(M)}_{q_M}$,
where $V_{j,p}$ is the representation space of $P_{\pm,0}$ and $H^{(i)}_{q_i}$ is the Fock space of $Q^{(i)}_{\pm,0}$.
   From (\ref{Jordan-C}) we have 
\begin{eqnarray*} 
Q_0^{(i)} = Q_0^{(M)} + \sum_{\mu=i}^{M-1}{\cal L}_{\mu} ,~~~~(M+1){\cal K}=MP_0+MQ_0^{(M)}+\sum_{\mu=1}^{M-1}\mu{\cal L}_{\mu}.
\end{eqnarray*}
This implies that for any irreducible representation of (\ref{plsu2}) defined by basis states $|j,n;p\ra \otimes \pd_{i=1}^M|q_i, m_i \ra$,
\begin{eqnarray*} 
m_i &=& m_M+q_M-q_i+\sum_{\mu=i}^{M-1}l_{\mu}, ~~~~i=1,\cdots,M-1,\n 
n+m_M&=&\frac{M+1}{M}\k -q_M-\frac{1}{M}\sum_{\mu=1}^{M-1}\mu l_{\mu}-\frac{p-j}{r},
\end{eqnarray*}
where $\k$ and $l_\mu$ denote the eigenvalues of central elements ${\cal K}$ and ${\cal L}_\mu$, respectively. 
It follows that $q_i\leq q_M+\sum_{\mu=i}^{M-1}l_{\mu}$  and 
\begin{eqnarray*} 
n+m_i= A_i, ~~~~~ A_i = \frac{M+1}{M}\k-\frac{1}{M}\sum_{\mu=1}^{M-1}\mu l_{\mu} +\sum_{\mu =i}^{M-1}l_{\mu} - \frac{p-j}{r}-q_i. 
\end{eqnarray*}
Clearly $A_i$ always take non-negative integer values, i.e. $A_i=0,1,\ldots$. Thus, the irreducible representation of (\ref{plsu2}) has basis states
\beqa
|j,n,p,\{q\},\{l\},\k \ra  &\equiv&  |j,n;p\ra \otimes \pd_{i=1}^M|q_i, m_i \ra \n
&=& \frac{J_+^{p+rn}|j,0\ra}{\sqrt{(p+rn)!(2j-p-rn)!}} \pd_{i=1}^M \frac{a_i^{\dg k_i(A_i+q_i-{k_i^{-2}}-n)}|0 \ra}
 {\sqrt{[k_i(A_i+q_i-{k_i^{-2}}-n)]!}}  \label{polybasis}
\eeqa  
where
\begin{eqnarray*} 
n=\left\{
\begin{array}{ll}
 0,1, \cdots, \textrm{min}\{ A_M, \frac{2j-p-\l}{r}\} & 
{\rm for}~M > 0 \\
0,1 \cdots \frac{2j-p-\l}{r} &
{\rm for}~ M=0
\end{array}
\right..
\end{eqnarray*}
The action of (\ref{Jordan-1}) on these states is given by
\begin{eqnarray*}
{\cal P}_0|j,n,p,\{q\},\{l\},\k \ra &=& \left(\frac{p-j}{r}+n-\k \right) |j,n,p,\{q\},\{l\},\k \ra, \\   
{\cal P}_+|j,n,p,\{q\},\{l\},\k \ra  &=&  \pd_{i=1}^r \sqrt{(p+i+rn)(2j-p-i+1-rn)}  \\
&&\times \pd_{i=1}^M\pd_{\mu=1}^{k_i}\left( A_i+q_i -\frac{(\mu-1)k_i+1}{k_i^2}-n\right)^{1/2}\\
&&\times |j,n+1,p,\{q\},\{l\},\k \ra, \\
{\cal P}_-|j,n,p,\{q\},\{l\},\k \ra &=&\pd_{i=1}^r \sqrt{(p-i+1+rn)(2j-p+i-rn)} \\ 
&&\times \pd_{i=1}^M\pd_{\mu=1}^{k_i}\left( A_i+q_i+ \frac{\mu k_i-1}{k_i^2}-n\right)^{1/2} 
|j,n-1,p,\{q\},\{l\},\k \ra.   
\end{eqnarray*} 
This gives an ${\cal N}+1$ dimensional representation of the polynomial algebra (\ref{plsu2}), where
\begin{eqnarray*}
{\cal N}=\left\{
\begin{array}{ll}
 \textrm{min}\{ A_M, \frac{2j-p-\l}{r}\} & {\rm for}~M > 0 \\
 \frac{2j-p-\l}{r} & {\rm for}~ M=0
\end{array}
\right..
\end{eqnarray*}

\sect{Differential operator realization}

The finite-dimensional irreducible representations in the proceeding section
can be realized by differential operators acting on ${\cal N}+1$-dimensional space of monomials with basis
 $\left\{1,z,z^2,...,z^{\cal N} \right \} $, by mapping 
the basis vectors (\ref{polybasis}) into monomials in $z$: 
\begin{eqnarray*}
|j,n,p,\{q\},\{l\}, \k\ra  \longrightarrow \frac{z^n}{\sqrt{(p+rn)!(2j-p-rn)!\pd_{i=1}^{M}[k_i(A_i+q_i-\frac{1}{k_i^2}-n)]!}}. 
\end{eqnarray*}
The corresponding single-variable differential operator realization of (\ref{Jordan-1}) in the monomial space takes the following form
\beqa 
{\cal P}_0 &=& z\frac{d}{dz}-\k+\frac{p-j}{r},\n
{\cal P}_+ &=& z \pd_{i=1}^r\left(2j-p-i+1-rz\frac{d}{dz}\right) \n 
&&\times  \pd_{i=1}^{M}\pd_{\nu=1}^{k_i}\sqrt{k_i}\left( A_i+q_i-\frac{(\nu-1)k_i+1}{k_i^2} -z\dz\right), \n
{\cal P}_- &=& \frac{z^{-1}}{\pd_{\mu=1}^M\sqrt{k_\mu}^{k_\mu}}\pd_{i=1}^r\left(rz\frac{d}{dz}+p-i+1 \right). \label{jordan-d}
\eeqa 
Note that ${\cal P}_-$ contains no singularities as $\pd_{i=1}^r(p-i+1)=0$ for all allowed $p$ values. 

We can thus equivalently represent Hamiltonian (\ref{bosonH2})
as the single-variable differential operator of order ${\cal M} \equiv$ max$\{r+\sum_{i=1}^M k_i, s\}$,
\beqa
H &=&\sum_{i=1}^{M} w_iN_i+ g' \left(rz\frac{d}{dz} -j+p\right)^s
    +gz^{-1}\pd_{i=1}^r\left(rz\frac{d}{dz}+p-i+1 \right)  \n 
 & & + gz  \pd_{i=1}^r\left(2j-p-i+1-rz\frac{d}{dz}\right) \n
 & & ~~\times  \pd_{i=1}^{M}\pd_{\nu=1}^{k_i}k_i\left( A_i+q_i-\frac{(\nu-1)k_i+1}{k_i^2}-z\frac{d}{dz}\right)\label{differentialH}
\eeqa
with
$$ 
N_i = k_i\left(-z\frac{d}{dz}+\frac{M+1}{M}\k -\frac{p-j}{r} +\sum_{\mu=i}^{M-1}l_\mu
  -\frac{1}{M}\sum_{\mu=1}^{M-1}\mu l_\mu \right) -\frac{1}{k_i}. 
$$

\sect{Exact solutions}

We will now solve for the Hamiltonian equation
\beq
H\psi(z)=E\,\psi(z)\label{hamilton-eqn}
\eeq
for the differential operator realizations by using the functional Bethe ansatz method \cite{Wiegmann94,Sasaki09,Sasaki01},
where $\psi(z)$ is the eigenfunction and $E$ is the corresponding eigenvalue. It is straightforward to verify
\beqa
 Hz^n&=& z^{n+1}g \pd_{i=1}^r\left(2j-p-i+1-rn\right) \pd_{i=1}^{M}\pd_{\nu=1}^{k_i}k_i\left( A_i+q_i-n-\frac{(\nu-1)k_i+1}{k_i^2}\right) \n 
   &&+~{\rm lower~order~terms},~~~~~~~~~  n\in {\bf Z}_+.\label{hzm=hzm+1}
\eeqa
This means that the differential operator ({\ref{differentialH}) is not exactly solvable.
However, it is quasi exactly solvable, since when $n={\cal N}$ the first term ($\sim z^{n+1}$) on the r.h.s. of (\ref{hzm=hzm+1}) is vanishing.
That is $H$ preserves an invariant polynomial subspace of degree ${\cal N}$,
\beqa 
H{\cal V} \subseteq  {\cal V}, ~~~~~{\cal V}= {\rm span} \{1,z,...,z^{\cal N}\} 
\eeqa 
Thus up to an overall factor, the eigenfunctions of (\ref{differentialH}) have the form
\beq
\psi(z)= \prod_{i=1}^{ \cal N}\left(z-\alpha_i\right),\label{w-function}
\eeq
where $\{\alpha_i\,|\,i=1,2,\cdots,{ \cal N}\}$ are roots of the polynomial
which will be specified by the associated Bethe ansatz equations (\ref{bethe-ansatz-eqn}) below.
We can rewrite the Hamiltonian (\ref{differentialH}) as
\beq
 H = \sum_{i=1}^{\cal M} P_i(z)\left(\frac{d}{dz} \right)^i + P_0(z),\label{expansionH}
\eeq
where $P_0(z)$ and $P_i(z)$ are polynomials in $z$ determined from the expansion of the products in (\ref{differentialH}). 

Dividing the Hamiltonian equation $H\psi= E\psi $ by $\psi$ gives us
\beq
E= \frac{H\psi}{\psi} = \sum_{i=1}^{\cal M}P_i(z)i!\sum_{n_1<n_2<...<n_i}^{\cal N}
\frac{1}{(z-\alpha_{n_1})...(z-\alpha_{n_i})}  +P_0(z).\label{e=hpsi/psi}
\eeq
The l.h.s. of (\ref{e=hpsi/psi}) is a constant, while the r.h.s is a meromorphic function in $z$ with at most simple poles.
For them to be equal, we need to eliminate all singularities on the r.h.s of (\ref{e=hpsi/psi}). We may achieve this by demanding
that the residues of the simple poles, $z=\alpha_i, i=1,2,..., { \cal N}$ should all vanish. This leads to the Bethe ansatz
equations for the roots $\{\alpha_i \}$ :
\beqa 
&&\sum_{i=2}^{\cal M}\;\sum_{n_1<n_2<...<n_{i-1} \ne \mu}^{ \cal N}\frac{P_i(\alpha_\mu)i!}{(\alpha_\mu-\alpha_{n_1})\cdots
(\alpha_\mu-\alpha_{n_{i-1}})}+P_1(\alpha_{\mu})=0, \n \n  
&& ~~~~~~~~~~~~~~\mu=1,2,\,\cdots,\,{ \cal N}.   \label{bethe-ansatz-eqn}
\eeqa
The wavefunction $\psi(z)$ (\ref{w-function}) becomes the eigenfunction of $H$ (\ref{differentialH}) in the space ${\cal V}$
provided that the roots $\{\alpha_i\}$ of the polynomial $\psi(z)$ (\ref{w-function}) are the solutions of (\ref{bethe-ansatz-eqn}).

Let us remark that the Bethe ansatz equation (\ref{bethe-ansatz-eqn}) is the necessary and sufficient condition for
the r.h.s. of (\ref{e=hpsi/psi}) to be independent of $z$. This is because when (\ref{bethe-ansatz-eqn}) is satisfied
the r.h.s. of (\ref{e=hpsi/psi}) is analytic everywhere in the complex plane (including points at infinity) and thus must be
a constant by Liouville's theorem.

To obtain the corresponding eigenvalue $E$, we consider the leading order expansion of $\psi(z)$,	
$$\psi(z)= z^{ \cal N}- z^{{ \cal N}-1}\sum_{i=1}^{ \cal N}\alpha_{i} +\cdots. $$
It can be directly shown  that the ${\cal P}_{\pm, 0}\psi(z)$ have the  expansions
\begin{eqnarray*}
{\cal P}_+\psi &=&  -z^{ \cal N}g \left( \pd_{i=1}^r\left(2j-p-i+1-r({ \cal N}-1)\right) \right. \\ 
& & \left.\times  \pd_{i=1}^{M}\pd_{\nu=1}^{k_i}\sqrt{k_i}\left( A_i+q_i-{\cal N}+1-\frac{(\nu-1)k_i+1}{k_i^2}\right)\right) 
   \sum_{i=1}^{\cal N}\a_i + \cdots, \\ 
{\cal P}_- \psi &\sim& z^{{\cal N}-1} +\cdots,\\ 
{\cal P}_0\psi &=& z^{\cal N}\left({\cal N} +\frac{p-j}{r} -\k\right) + \cdots. 
\end{eqnarray*}
Substituting these expressions into the Hamiltonian equation (\ref{hamilton-eqn}) and equating the $z^{ \cal N}$ terms, we arrive at
\beqa
E &=& \sum_{i=1}^Mw_i\left(k_i\left(\frac{M+1}{M}\k-\frac{p-j}{r} -{\cal N}-\frac{1}{M}\sum_{\mu=1}^{M-1}l_{\mu}
   +\sum_{\mu=i}^{M-1}l_{\mu}\right)-\frac{1}{k_i}\right)\n
& & + g'\left( r{\cal N}-j+p\right)^{s}   - g\left[ \pd_{i=1}^r\left(2j-p-i+1-r({ \cal N}-1)\right) \right. \n 
& & \left.\times  \pd_{i=1}^{M}\pd_{\nu=1}^{k_i}k_i\left( A_i+q_i-{\cal N}+1-\frac{(\nu-1)k_i+1}{k_i^2}\right)\right] \sum_{i=1}^{\cal N}\a_i,
\label{energy-generalH}
\eeqa
where $\{\alpha_i\}$ satisfy the Bethe ansatz equations (\ref{bethe-ansatz-eqn}). This gives
the eigenvalue of the  Hamiltonian (\ref{bosonH1}) with the corresponding eigenfunction $\psi(z)$ (\ref{w-function}).

\sect{ Explicit examples }

In this section we give explicit results on the Bethe ansatz equations and energy eigenvalues
of the Hamiltonian (\ref{bosonH1}) for special cases which correspond to some established models frequently studied 
in the field of atomic and molecular physics, condensed matter, nuclear physics and quantum optics. 
\vskip.1in
\noindent \textbf{A. Two-site Bose-Hubbard model}
\vskip.1in
This model corresponds to the special case with $M=0, r=1, s=2$ and its Hamiltonian takes the simple form
\beq
H= g'J_0^2 +  g\left(J_+ +J_- \right).\label{h11}
\eeq
This model has been widely employed in the context of Josephson-coupled Bose-Einstein condensates via the realization of $J_{\pm,0}$ in terms of two bosons,
$J_+=b_1^\dg b_2,~J_-=b_1b_2^\dg,~J_0=\frac{1}{2}(b_1^\dg b_1-b_2^\dg b_2)$  
(see e.g. \cite{Leggett01} and references therein). Exact solutions of the model in terms of algebraic Bethe ansatz methods were first studied in \cite{Enolskii}.
From the general results in the preceding section, in this case we have $\k=0, p=0$, 
and ${\cal N}=2j$. Thus (\ref{h11}) takes the form
\begin{eqnarray*}
H= P_2(z)\frac{d^2}{dz^2}+P_1(z) \frac{d}{dz}+P_0(z),
\end{eqnarray*}
where
\begin{eqnarray*}
P_2(z)&=& g'z^2, \\
P_1(z)&=& g'z(1-2j)+g(1+z^2), \\ 
P_0(z) &=& g'j^2-2jzg.
\end{eqnarray*}
The Bethe ansatz equations are given by
\begin{eqnarray*}
\sum_{i \ne {\mu}}^{2j}\frac{2}{\a_i-\a_{\mu}} = - \frac{\a_{\mu}g'(1-2j)+g(1+\a_{\mu}^2) }{ g'\alpha_{\mu}^2}, ~~~~~\mu=1,2,\cdots,2j
\end{eqnarray*}
and the energy eigenvalues are
\begin{eqnarray*}
E &=& g'j^2 -g\sum_{i=1}^{2j}\a_i.
\end{eqnarray*}
This exact solution is equivalent to a case described in \cite{LinksHibberd06}.

\vskip.2in
\noindent\textbf{B. Lipkin-Meshkov-Glick model} 
\vskip.1in
This model is the special case corresponding to $M=0, r=2,s=1$. The Hamiltonian is given by \cite{Lipkin65}
\beq
H= g'J_0 +  g\left(J_+^2+J_-^2 \right) \label{LMG-H}
\eeq
and continues to be studied extensively (see e.g. \cite{Ribeiro07} and references therein). 
Exact solution via the algebraic Bethe ansatz method is discussed in
\cite{Pan99,Morita06}. Specializing the general results of the preceding section to this case,
we have $p=0,1$, $\k=0$ and ${\cal N}=\frac{2j-p-\l}{2}$ with $\l=0,1$ so that ${\cal N}$ is a non-negative integer.
The differential operator representation of the Hamiltonian (\ref{LMG-H}) is thus
\begin{eqnarray*}
H= P_2(z) \frac{d^2}{dz^2}+ P_1(z) \dz +P_0(z)
\end{eqnarray*}
where
\begin{eqnarray*}
P_2(z)&=& 4gz^{3}+ 4gz, \\
P_1(z)&=&  g(6+4p-8j){z}^{2}+ 2g'z+ g( 2+4p ),\\
P_0(z) &=& gz(2j-p)(2j-p-1) +g'(p-j).
\end{eqnarray*}
The Bethe ansatz equations are given by
$$
\sum_{i \ne \mu}^{ \frac{2j-p-\l}{2}} {2\over \a_i - \a_{\mu}}= - \frac{g(3+2p-4)\a_{\mu}^2+g'\a_{\mu}+g(1+2p)}{2g(\a_{\mu}^{3}+ \a_{\mu}) },
$$
\begin{eqnarray*}
\mu=1,2,\cdots, \frac{2j-p-\l}{2}
\end{eqnarray*}
and the energy eigenvalues are 
\begin{eqnarray*}
E &=& g'\left(j-\l \right) - g (\l+1)(\l+2)\sum_{i=1}^{\frac{2j-p-\l}{2}}\a_i.
\end{eqnarray*}

\vskip.2in
\noindent \textbf{C. Molecular asymmetric rigid rotor}
\vskip.1in
Up to an additive constant, this model corresponds to the special case with $M=0,r=s=2$. This is shown as follows.
The Hamiltonian of the rigid rotor has the following form in terms of the $su(2)$ generators $J_x, J_y$ and $J_z$ \cite{King43}: 
\begin{eqnarray*}
H= a J_x^2 + b J_y^2+c J_z^2  
\end{eqnarray*}
where  $a, b, c$ are constants. The model has previously been discussed in \cite{Jarvis08} as a Hamiltonian 
which is solvable by algebraic Bethe ansatz methods. The Hamiltonian can be rewritten as 
\beqa 
H = \frac{2c-a-b}{2} J_0^2 + \frac{a-b}{4}(J_+^2 + J_-^2) + \frac{a+b}{2} C \label{rot-ham2}
\eeqa 
where $C$ is the Casimir element of $su(2)$. This shows that the molecular asymmetric rigid rotor is indeed 
a special case of  (\ref{bosonH1}). Note that the Hamiltonian (\ref{rot-ham2}) of the rigid rotor 
almost has the same form as that of the Lipkin-Meshkov-Glick model. To our knowledge, this connection has not been noted previously.

Specializing the general results in the preceding section to this case,
we have $p=0,1$, $\k=0$ and ${\cal N}=\frac{2j-p-\l}{2}$, where $\l=0,1$ as in the case of the Lipkin-Meshkov-Glick  model.
The differential operator representation of the Hamiltonian (\ref{rot-ham2}) is thus
\begin{eqnarray*}
H= P_2(z) \frac{d^2}{dz^2}+ P_1(z) \dz +P_0(z)
\end{eqnarray*}
where
\begin{eqnarray*}
P_2(z)&=& (a-b)z^{3}+ 2(2c-a-b)z^2+ (a-b)z, \\
P_1(z)&=&  \frac{a-b}{2}(3+2p-4j){z}^{2}+ 2(2c-a-b)(1+p-j)z+ \frac{a-b}{2}( 1+2p ),\\
P_0(z) &=& \frac{a-b}{4}(2j-p)(2j-p-1)z +\frac{2c-a-b}{2}(p-j)^2 + \frac{a+b}{2}j(j+1).
\end{eqnarray*}
The Bethe ansatz equations are given by
$$
\sum_{i \ne \mu}^{ \frac{2j-p-\l}{2}} {2\over \a_i - \a_{\mu}}
= - \frac{(a-b)(3+2p-4)\a_{\mu}^2+4(2c-a-b)(1+p-j)\a_{\mu}+(a-b)(1+2p)}{2(a-b)(\a_{\mu}^{3}+ \a_{\mu}) +4(2c-a-b)\a_{\mu}^2},
$$
\begin{eqnarray*}
\mu=1,2,\cdots, \frac{2j-p-\l}{2}
\end{eqnarray*}
and the energy eigenvalues are
\begin{eqnarray*}
E &=&\frac{2c-a-b}{2}\left(j-\l \right)^2  + \frac{a+b}{2} j(j+1) - \frac{a-b}{4} (\l+1)(\l+2)\sum_{i=1}^{\frac{2j-p-\l}{2}}\a_i .
\end{eqnarray*}

\newpage
\noindent \textbf{D. Tavis-Cummings model} 
\vskip.1in
This model corresponds to the special case when $M=r=s=k_1=1$. The Hamiltonian is given by
\begin{eqnarray*}
H= w_1N_1 + g'J_0 +  g\left(J_+a_1 +J_-a_1^{\dg}\right).
\end{eqnarray*}
This is one of the widely studied models in quantum optics and had been exactly solved via the 
algebraic Bethe ansatz approach \cite{Bogoliubov96,Amico05}. Applying the results in the preceding section gives $q_1=1$, $p=0$ and
${\cal N}=\textrm{min}\{2\k  +j-1 , 2j\}$. The differential operator representation of the Hamiltonian is 
\begin{eqnarray*}
H= P_2(z) \frac{d^2}{dz^2}+ P_1(z) \dz +P_0(z),
\end{eqnarray*}
where
\begin{eqnarray*}
P_2(z)&=& gz^3,\\
P_1(z)&=&  -g(3j+2\k-2)z^2 +(g'-w_1)z+g,\\
P_0(z)&=&  w_1(2\k+j-1) -g'j+2gjz(2\k+j-1).
\end{eqnarray*}
The Bethe ansatz equations read
$$
\sum_{i \ne \mu}^{ \cal N} \frac{2}{\a_i- \a_\mu }= \frac{g(3j+2\k-2)\a_\mu^2-(g'-w)\a_\mu -g}{g\a_\mu^3 }, ~~~~\mu=1,2,\cdots, {\cal N}
$$
and the energy eigenvalues are
\begin{eqnarray*}
E = w_1\left(2\k +j -{\cal N}-1\right) + g'\left( {\cal N}-j\right)
- g \left(2j-{ \cal N}+1\right) \left( 2\k+j -{\cal N}\right)\sum_{i=1}^{\cal N}\a_i,
\end{eqnarray*}
where ${\cal N}=\textrm{min}\{2\k  +j-1 ,2j \}$.

\vskip.2in
\noindent \textbf{E. Two-mode generalized Tavis-Cummings model}
\vskip.1in
Finally, we consider the case when $M=2, ~~r=s=k_1=k_2=1$. This gives the Hamiltonian
\begin{eqnarray*}
H= w_1N_1+w_2N_2 + g'J_0 +  g\left(J_-a_1^{\dg}a_2^{\dg}+J_+a_1a_2 \right),
\end{eqnarray*}
which belongs to the class of $su(1,1)$ generalized Tavis-Cummings model discussed in \cite{Rybin98}. 
Applying the results in the preceding section gives $q_1=q_2=1$, $p=0$ and
${\cal N}=\textrm{min}\{\frac{3\k-l_1}{2} -1 +j , {2j}\}$. The differential operator representation of the Hamiltonian thus reads
\begin{eqnarray*}
H= P_3(z)\frac{d^3}{dz^3}+P_2(z) \frac{d^2}{dz^2}+ P_1(z) \dz +P_0(z),
\end{eqnarray*}
where
\begin{eqnarray*}
P_3(z)&=& -gz^4, \\ 
P_2(z)&=& g(3\k+4j-5)z^3,\\
P_1(z)&=& Az^2+ (g'-w_1-w_2)z+g,\\
P_0(z)&=&zB + F
\end{eqnarray*}
with
\begin{eqnarray*}
A &=& g\left(-9j\k+10j+6\k+\frac{l_1^2}{4}-5j^2-4-\frac{9}{4}\k^2\right), \\ 
B &=& g\left( \frac{9j\k}{2}+6j^2\k-6j\k+2j-\frac{jl_1^2}{2}+2j^3-4j^2 \right), \\ 
F &=& (w_1+w_2)\left(\frac{3\k}{2}-1+j  \right)+\frac{l_1}{2}(w_1-w_2) -g'j. 
\end{eqnarray*}
The Bethe ansatz equations assume the form
$$
\sum_{\mu <\nu \ne {\beta} }^{{\cal N}} \frac{6g\a_ {\beta}^4}{(\a_ {\beta}-\a_{\mu})(\a_ {\beta}-\a_{\nu})} 
-\sum_{i \ne  {\beta}}^{\cal N}\frac{2g(3\k+4j-5)\a_ {\beta}^3}{\a_ {\beta}-\a_i} = A\a_ {\beta}^2+ (g'-w_1-w_2)\a_ {\beta}+g, \n 
$$ 
\begin{eqnarray*}
 {\beta}=1,2,\cdots,{\cal N}
\end{eqnarray*}
and the energy eigenvalues are
\begin{eqnarray*}
E &=& (w_1+w_2)\left(\frac{3\k}{2} -1+j -{\cal N}\right)+ \frac{l_1}{2}(w_1-w_2) + g'\left( {\cal N}-j\right) \n 
&& - g( 2j-{\cal N}+1)\left[\left( \frac{3\k}{2}+j-{\cal N}\right)^2-\frac{l_1^2}{4} \right ] \sum_{i=1}^{\cal N}\a_i,
\end{eqnarray*}
where ${\cal N}=\textrm{min}\{\frac{3\k-l_1}{2} -1 +j , {2j}\}$.

\sect{Discussions}
We have derived the exact solutions of a family of Hamiltonians with the following general form, 
\beqa
H= F({\cal Q}_0)+g({\cal Q}_+ +{\cal Q}_-) 
\eeqa 
whereby ${\cal Q}_{\pm,0}$ are particular polynomial deformations of the $sl(2)$ Lie algebra and $F({\cal Q}_0)$ is a polynomial function of ${\cal Q}_0$ with
real coefficients. 
We have seen that via the differential operator realization of these algebras, the block diagonal sectors of the Hamiltonians can be realized as higher 
order quasi-exactly solvable differential operators. The eigenvalues of the Hamiltonians in these sectors have been obtained 
via the functional Bethe ansatz approach.

Specific cases of the general Hamiltonian have previously been solved via the algebraic Bethe ansatz approach \cite{Jarvis08,Enolskii,Pan99,Morita06,Bogoliubov96,Amico05} as mentioned earlier.
 Comparing both methods, it appears that the functional Bethe ansatz approach has some  
advantage over the algebraic Bethe ansatz by requiring less algebraic machinery. This advantage manifests itself in the fact that we have been able to give a unified
solution for (\ref{bosonH1}) through (\ref{bethe-ansatz-eqn},\ref{energy-generalH}). Such a unified solution presently appears beyond the limits of algebraic Bethe ansatz approaches which treat the models on a case-by-case basis. It would therefore be interesting to see whether other classes of exactly solvable models 
can be easily handled by the functional Bethe ansatz approach.   

One avenue for further work would be to generalize the functional Bethe ansatz approach to solve for other classes of Hamiltonians,
such as $q$-deformed versions of the  models discussed above.
It would also be worthwhile to explore the role of polynomial algebra structures in connections between higher order ODEs and integrable models 
i.e. the ODE/IM correspondence \cite{Dorey99}.

\vskip.2in
\noindent {\bf Acknowledgments:} This work was supported by the Australian Research Council.

\bebb{99}

\bbit{Karassiov94}
V.P. Karassiov and A. Klimov, Phys. Lett. A {\bf 191}, 117 (1994).

\bbit{Karassiov92-00}
V.P. Karassiov, J, Sov. Laser Res. {\bf 13}, 188 (1992); J. Phys. A: Math. Gen. {\bf 27}, 153 (1994);
J. Russian Laser Res. {\bf 21}, 370 (2000).

\bbit{Karassiov02}
V.P. Karassiov, A.A. Gusev and S.I. Vinitsky, Phys. Lett. A {\bf 295}, 247 (2002).

\bbit{Yuan10}
Y.-H. Lee, W.-L. Yang and Y.-Z. Zhang, J. Phys. A: Math. Theor. {\bf 43}, 185204 (2010);
J. Phys. A: Math. Theor. {\bf 43}, 375211 (2010). 

\bbit{Leggett01}
A.J. Leggett, Rev. Mod. Phys. {\bf 73}, 307 (2001).

\bbit{Lipkin65}
H.J. Lipkin, N. Meshkov and A.J. Glick, Nucl. Phys. {\bf 62}, 188 (1965).

\bbit{King43} G.W. King, R.M. Hainer, and P.C. Cross, J. Chem. Phys. {\bf 11}, 27 (1943).

\bbit{Tavis68}
M Tavis and F.W. Cummings, Phys. Rev. {\bf 170}, 379 (1968).

\bbit{Hepp73}
K. Hepp and E.H. Lieb, Ann. Phys. {\bf 76}, 360 (1973).

\bbit{Rybin98}
A. Rybin, G. Kostelewicz, J. Timonen and N.M. Bogoliubov, J. Phys. A: Math. Gen. {\bf 31}, 4705 (1998).

\bbit{Turbiner88}
A. Turbiner, Comm. Math. Phys. {\bf 118}, 467 (1988).

\bbit{Ushveridze94}
A.G. Ushveridze, Quasi-exactly solvable models in quantum mechanics, Institute of Physics
Publishing, Bristol, 1994.

\bbit{Gonzarez93}
A. Gonz\'arez-L\'opez, N. Kamran and P. Olver, Commun. Math. Phys. {\bf 153}, 117 (1993).

\bbit{Wiegmann94}
P.B. Wiegmann and A.V. Zabrodin, Phys. Rev. Lett. {\bf 72}, 1890 (1994);
 Nucl. Phys.  B {\bf 451}, 699 (1995).

\bbit{Sasaki09}
R. Sasaki, W.-L. Yang and Y.-Z. Zhang, SIGMA {\bf 5}, 104 (2009).

\bbit{Sasaki01} R. Sasaki and K. Takasaki, J. Phys. A: Math. Gen. {\bf 34}, 9533 (2001).

\bbit{Enolskii} V.Z. Enol'skii, M. Salerno, N.A. Kostov and A.C. Scott, Phys. Scripta {\bf 43}, 229 (1991); \\
V.Z. Enol'skii, M. Salerno, A.C. Scott  and J.C. Eilbeck, Physica D {\bf 59}, 1 (1992); \\
V.Z. Enol'skii, V.B. Kuznetsov and M. Salerno, Physica D {\bf 68}, 138 (1993).

\bbit{LinksHibberd06} J. Links and K. Hibberd, SIGMA {\bf 2}, 094 (2006).

\bbit{Ribeiro07}
P. Ribeiro, J. Vidal and R. Mosseri, Phys. Rev. Lett. {\bf 99}, 050402 (2007); Phys. Rev. E {\bf 78},  021106 (2008).

\bbit{Pan99}
F. Pan and J.P. Draayer, Phys. Lett. B {\bf 451}, 1 (1999).


\bbit{Morita06}
H. Morita, H. Ohnishi, J. da Providenica and S. Nishiyama, Nucl. Phys. B {\bf 737}, 337 (2006).

\bbit{Jarvis08} 
P.D. Jarvis and L.A. Yates, Mol. Phys. {\bf 106}, 955 (2008).

\bbit{Bogoliubov96}
N.M. Bogoliubov, R.K. Bullough and J. Timonen, J. Phys. A: Math. Gen. {\bf 29}, 6305 (1996).

\bbit{Amico05}
L. Amico and K. Hikami, Eur Phys. J. {\bf B 43}, 387 (2005);\\
L. Amico, H. Frahm, A. Osterloh and G.A.P. Ribeiro, Nucl. Phys. {\bf B 787}, 283 (2007).


\bbit{Dorey99}
P. Dorey and R Tateo, J. Phys. A: Math. Gen. {\bf 32}, L419 (1999).

\eebb

\end{document}